\documentclass[conference]{IEEEtran}
\usepackage{cite}
\usepackage{amsmath,amssymb,amsfonts}
\usepackage{algorithmic}
\usepackage{graphicx}
\usepackage{textcomp}
\usepackage{xcolor}
\def\BibTeX{{\rm B\kern-.05em{\sc i\kern-.025em b}\kern-.08em
    T\kern-.1667em\lower.7ex\hbox{E}\kern-.125emX}}
    
\begin{document}

\title{Preamble-Based Packet Detection in Wi-Fi:\\ A Deep Learning Approach\\
}

\author{
\IEEEauthorblockN{Vukan Ninkovic, Dejan Vukobratovic}
\IEEEauthorblockA{
\textit{Faculty of Technical Sciences} \\
\textit{University of Novi Sad}\\
Novi Sad, Serbia \\
email: \{ninkovic,dejanv\}@uns.ac.rs
}
\and
\IEEEauthorblockN{Aleksandar Valka, Dejan Dumic}
\IEEEauthorblockA{
\textit{Methods2Business} \\
Novi Sad, Serbia \\
email: \{aleksandar,dejan\}@methods2business.com}
}

\maketitle

\begin{abstract}
Wi-Fi systems based on the family of IEEE 802.11 standards that operate in unlicenced bands are the most popular wireless interfaces that use Listen Before Talk (LBT) methodology for channel access. Distinctive feature of majority of LBT-based systems is that the transmitters use preambles that precede the data to allow the receivers to acquire initial signal detection and synchronization. The first digital processing step at the receiver applied over the incoming discrete-time complex-baseband samples after analog-to-digital conversion is the packet detection step, i.e., the detection of the initial samples of each of the frames arriving within the incoming stream. Since the preambles usually contain repetitions of training symbols with good correlation properties, conventional digital receivers apply correlation-based methods for packet detection. Following the recent interest in data-based deep learning (DL) methods for physical layer signal processing, in this paper, we challenge the conventional methods with DL-based approach for Wi-Fi packet detection. Using one-dimensional Convolutional Neural Networks (1D-CNN), we present a detailed complexity vs performance analysis and comparison between conventional and DL-based Wi-Fi packet detection approaches.     

\end{abstract}

\begin{IEEEkeywords} 

Deep Learning, Packet Detection, IEEE 802.11
\end{IEEEkeywords}

\section{Introduction}

In order to ensure fairness, wireless systems operating in unlicenced bands share a common channel access approach based on Listen Before Talk (LBT) methodology. Among such systems, by far the most pervasive are the Wi-Fi systems based on the IEEE 802.11 standards that apply carrier-sense multiple access with collision avoidance (CSMA/CA) method \cite{b4}. Due to spectrum scarcity, cellular systems are also introducing usage of LBT methods in unlicenced bands, for example, in unlicenced version of 4G standard called LTE-U \cite{lte-u}.

Common approach in majority of LBT systems is that the transmitters send preambles prepended to the data packets in order to ensure that the receivers detect signal and acquire initial synchronization. Such preambles are usually designed as sequences of symbols with good correlation properties, allowing the signal processing algorithms at the receiving end to identify packet start samples \cite{b4, schlegel_2006}. In particular, at the receiver side, after the signal is converted into a stream of discrete-time complex baseband samples, the first digital receiver block represents the packet detection algorithm. Conventional digital receivers apply correlation-based methods to detect the initial samples of each frame arriving within the incoming stream \cite{b3,b10}.  

In recent years, data-based approaches relying on deep learning (DL) demonstrated excellent performance when applied for signal processing tasks at the wireless receiver side \cite{OShea_2017, Qin_2019}.  Different studies focused on different physical layer  (PHY) signal processing tasks,  ranging across signal  detection \cite{Karra_2017}, channel  estimation \cite{Ye_2017}  and error correction coding \cite{Nachmani_2018}. In a companion paper, we investigated DL-based signal detection and carrier frequency offset (CFO) estimation in IEEE 802.11 systems, evaluating various DL architectures for packet detection and CFO estimation against conventional algorithms \cite{Ninkovic_2020}. In this paper, we focus on the packet detection part and take this analysis a step further, by providing detailed complexity vs performance evaluation and comparison between conventional and DL-based Wi-Fi packet detection. Using one-dimensional Convolutional Neural Networks (1D-CNN) whose excellent performance for sequence detection is demonstrated in \cite{Karra_2017, Ninkovic_2020}, we perform fine-grained evaluation and comparison of 1D-CNN architectures of different parameters, against the conventional correlation-based packet detector. Our results demonstrate that 1D-CNN architectures may outperform conventional methods, both in the performance and computational complexity, while maintaining robustness at low signal-to-noise ratio (SNR).

The paper is organized as follows. In Sec. II, we provide background information and present the system model. In Sec. III, we review conventional correlation-based packet detection, and present details on 1D-CNN-based packet detection. Detailed complexity vs performance analysis of both methods is given in Sec. IV. The paper is concluded in Sec. V.     

\section{Background and System Model}

\begin{figure}[t]
 \centerline{\includegraphics[width=.9\columnwidth, height=0.75in]{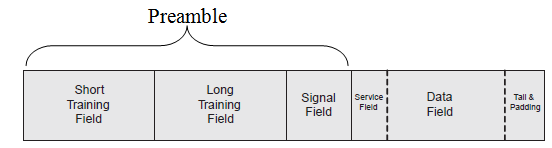}}
  \caption{IEEE 802.11 frame structure.}
  \label{fig:OFDM_frame_structure}
\end{figure}

In this paper, we focus on the IEEE 802.11 technology based on the orthogonal frequency division multiplexing (OFDM). Therein, the preamble based on repeated patterns of symbols with good correlation properties is prepended to data symbols for initial synchronization and/or channel estimation (Fig. \ref{fig:OFDM_frame_structure}). The initial synchronization procedure consists of: i) the packet detection, ii) sampling time offset (TO) estimation, and iii) the CFO estimation \cite{b5}. 

Without loss of generality, and for the purpose of implementation and evaluation, we focus on the IEEE 802.11ah (\textit{Wi-Fi HaLow}) standard for Wi-Fi Internet of Things (IoT) \cite{b6}. For 1 MHz packet structure, the 802.11ah preamble comprises 14 OFDM symbols, where each OFDM symbol contains $N=32$ subcarriers spaced at $\Delta f = 31.25$ kHz in the frequency domain. In the time domain, each OFDM symbol is extended by a cyclic prefix of $8 \mu s$ duration, resulting in $40 \mu s$ OFDM symbol. The preamble is divided into three fields:  

\textbf{Short Training Field (STF):} STF consists of 4 OFDM symbols which, after IDFT, represent 10 repetitions of the same $16 \mu s$-long short training symbol (STS) in the time domain. STS is of good correlation properties and is suitable for coarse timing synchronization and coarse CFO estimation.

\textbf{Long Training Field 1 (LTF1):} LTF1 also contains 4 OFDM symbols of $160 \mu s$ duration. Two repetitions of the same long training symbol (LTS) enable fine timing synchronization, CFO estimation and channel estimation.

\textbf{Signal Field (SIG):} contains packet information to configure the receiver, while \textbf{Long Training Field 2 (LTF2)} is used for MIMO channel estimation, while herein, we focus on single-antenna (SISO) transmission.

\begin{figure}[htbp]
 \centerline{\includegraphics[width=3in, height=2.2in]{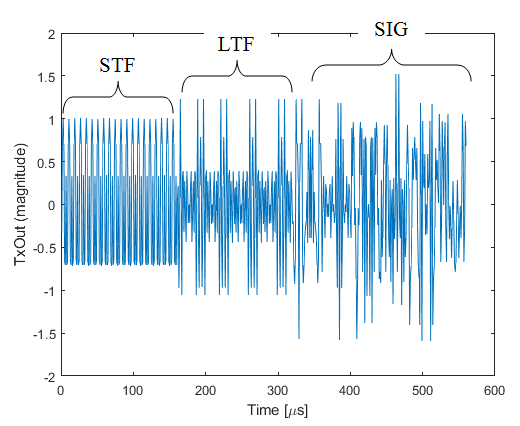}}
  \caption{802.11ah NDP packet transmit waveform.}
  \label{fig:NDP_packet}
\end{figure}

Fig. \ref{fig:NDP_packet} shows magnitude of discrete-time complex baseband 802.11ah preamble of total duration $560 \mu s$. Since our focus is on the detection of initial packet sample, in order to reduce the simulation load, we use 802.11ah Null Data Packets (NDP) that contain only the preamble without the data field \cite{b5}.

\vspace{-0.2cm}

Before transmission, the time-domain samples $\pmb{x}$ are over-sampled and filtered, resulting in the over-sampled signal samples $\pmb{x}_{os}$. Focusing on the discrete-time complex-baseband model, the channel is represented via its equivalent discrete-time impulse response $\pmb{h}$. After adding complex Additive White Gaussian Noise (AWGN) $\pmb{w}$ samples, the received signal samples are modelled as: 
\begin{equation}
\pmb{y}_{os} = \pmb{x}_{os} \circledast \pmb{h} + \pmb{w},
\end{equation}
where $\circledast$ represents the circular convolution. Besides the channel impairments and the noise, the received signal ($\pmb{y}_{os}$) is affected by the sampling TO $\epsilon$ and the CFO $\Delta=f_{off}/\Delta f$ which needs to be estimated and corrected \cite{b3}. At the receiver, after the signal undergoes reverse pulse-shape filtering and down-sampling, the received samples $\pmb{y}$ are forwarded to the packet detection, which we detail in the next section.

\section{Preamble-Based Packet Detection}

\subsection{Conventional Packet Detection Methods}

Conventional algorithms for packet detection, which are nowadays widely used, use repetitive preamble structure through complex correlation between two subsequently received training symbols. If we suppose that the number of complex samples in one training symbol is $L$, such complex correlation can be expressed as:

\begin{equation}
\Lambda_{\tau} = \sum_{i=0}^{L-1} \ y^{*}_{\tau+i} y_{\tau+i+L}
\end{equation}

 In \cite{b3} and \cite{b10}, authors proposed packet detection algorithm which relies on assumption that the channel effects will be annulled  if the conjugated sample from one training symbol is multiplied by corresponding sample from adjacent training symbol. Consequently, products of these sample pairs at the start of the frame will have approximately the same phase, thus the magnitude of their sum will be a large value. In order to reduce the complexity of the algorithm, they introduced a window of $2L$ samples which slides along time $\tau$ as the receiver searches for the first training symbol, i.e., the packet start sample $\tau_S$. Timing metric used for packet detection is: 

\begin{equation}
M(\tau) = \frac{|\Lambda_{\tau}|^2}{P_{\tau}^2},
\end{equation}
where $P_{\tau}$ is the sum of the powers of $L$ subsequent samples:

\begin{equation}
P_{\tau} = \sum_{i=0}^{L-1} \ |y_{\tau+i}|^2
\end{equation}

From the timing metric $M(\tau)$, one may find the initial packet sample by finding the sample that maximizes $M(\tau)$. In addition, except finding the maximum sample-point, observing the points to the left and right in the time domain which are at the 90\%\ of the maximum, and averaging these two $90\%$-time samples, may result in more accurate timing estimation. A threshold which triggers the above algorithm should be chosen in a way that the algorithm minimizes the probability of miss detection while controlling for the probability of false alarm.

Packet detection in IEEE 802.11 is usually separated into two steps: coarse and fine synchronization, where the main principles from conventional algorithms are reused and adapted to the specific system requirements. The coarse packet detection, denoted as $\hat{\tau}_{S}$, may follow \cite{b3} (Eq. 3), setting $L=80$ samples (one half of STF duration):

\begin{equation}
\begin{gathered}
    \hat{\tau}_{S}=\arg\max_{\tau}\frac{|\Lambda_{\tau}|^2}{(P_{\tau})^2}\\
    =\arg\max_{\tau}(\frac{|\sum_{n=\tau}^{\tau+L_{S}-l_{S}} y^*_{n} y_{n+l_{S}}|^2}{(\sum_{n=\tau}^{\tau+L_{S}-l_{S}} |y_{n+l_{S}}|^2)^2}),
\end{gathered} 
\end{equation}
where $l_{S}$ is the STS sample-length and $L_{S}$ represents sample-lengths of STF field. After $\hat{\tau}_{S}$ is obtained, we can extract the whole preamble because the peaks from the correlation between a single long training symbol and the entire preamble are used to derive a more accurate time estimation \cite{b4}. 

\subsection{Deep-Learning Based Packet Detection}

\textbf{Convolutional Neural Networks for Packet Detection:} Motivated by the initial results in our companion paper \cite{Ninkovic_2020}, and recent investigation performed in \cite{Karra_2017}, we consider Wi-Fi packet detection using one-dimensional convolutional neural networks (1D-CNN) which provide excellent results in processing time series data.

\begin{figure}[htbp]
 \centerline{\includegraphics[width=0.85\columnwidth, height=2.4in]{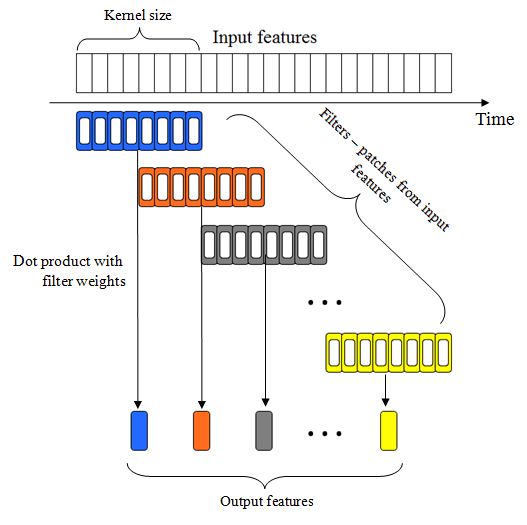}}
  \caption{Structure of 1D convolution layer.}
  \label{fig:1D-CNN}
\end{figure}

CNNs are DL architectures that achieved outstanding results in computer vision and image classification problems, due to their ability to extract features from local input patches through the application of relevant filters. CNNs can effectively learn the hierarchical features to construct a final feature set of a high level abstraction, which are then used to form more complex patterns within higher layers \cite{b20}. The same ideas can be applied to 1D-sequences of data, where 1D-CNNs are proven to be effective in deriving features from fixed-length segments of the data set. This characteristic of the 1D-CNN, together with the fact that the 1D convolution layers are translation invariant (a pattern learned at a certain position in the signal can be latter recognized at a different position), makes this architecture suitable for packet detection tasks.

Two types of layers are applied in compact 1D-CNNs: i) 1D-CNN layer, where 1D convolution occurs, and ii) Fully Connected (FC) layer. Each hidden CNN layer performs a sequence of convolutions, whose sum is passed through the activation function \cite{b21}. The main advantage of 1D-CNN represents fusing feature extraction and classification operations into a single process that can be optimized to maximize the network performance, because CNN layers process the raw 1D data and extract features used by FC layers for prediction tasks (Fig. \ref{fig:1D-CNN}). As a consequence, low computational complexity is provided and if compared to 2D-CNNs, 1D-CNN can use larger filter and convolution window sizes since the only expensive operation is a sequence of 1D convolutions.

\textbf{Data Set and Training Procedure:} The packet detection problem can be formulated as a regression problem, where CNN needs to learn a mapping between the input signal and the output value representing the packet start instant while distinguishing from the noise. We suppose that CNN-based packet detection operates over the consecutive fixed-length blocks $|\pmb{y}|$ of the received signal amplitude samples:

\begin{equation}
    \hat{\tau_{S}} = f(|\pmb{y}|),
\end{equation}
after the received signal is down-sampled and filtered. The data set consists of ($|\pmb{y}|, \tau_{S}$) pairs, where $\tau_{S}$ indicates a packet start sample within the block. Within the data set, we included about $50 \%$ of the blocks that do not contain a packet start instance, tagged with the value of $\tau_{S}=-1$. Among such blocks, roughly half contain only noise samples, while other half contain intermediate or tail-parts of NDP packets. For data set blocks containing packet start instants $\tau_{S}$, its value is set uniformly at random among the input block samples. Data sets are created for input blocks $|\pmb{y}|$ of lengths: 40, 80, 160, 320, 800, 1600 samples, where the number of received blocks in each data set is 50000. From the data set, $70\%$ records are used for training, $15\%$ for validation and $15\%$ for testing.

%\vspace{-0.3cm}
%\begin{table}[tbhp]
%\caption{ Input block size vs Data set size}
%\begin{center}
%\begin{tabular}{|c|c|}
%\hline
%Input block size&Data set size [number of received blocks] \\
%\hline 
%25&80000\\
%50&66000\\
%100&58000\\
%200&52000\\
%400&46000\\
%800&40000\\
%1600&30000\\
%\hline
%\end{tabular}
%\label{table_1}
%\end{center}
%\end{table}

Regardless of the input block size, all packets are simulated under the same conditions using the standard-compliant IEEE 802.11ah physical layer simulator. In order to examine estimator robustness to varying signal-to-noise-ratio (SNR), SNR values are uniformly and randomly selected from range [0 dB, 25 dB]. During the simulations, indoor multipath fading channel - model B \cite{b17} is applied.

To train DNN models, the mean-squared error (MSE) loss: $L_{MSE}(\tau_{S}, \hat{\tau}_{S}) = \sum_{i}(\tau_{S_i}-\hat{\tau}_{S_i})^2$ is minimized, since we observed that the  algorithm achieved better performances as compared to the mean-absolute error (MAE) and Huber loss functions. The training set is separated into mini-batches with size 80, and 400 epochs are sufficient for the loss function convergence. In order to optimize network parameters, stochastic gradient descent (SGD) algorithm with ADAM optimizer at the learning rate $\alpha = 0.001 $, $\beta_{1} = 0.9$ and $\beta_2 = 0.999$ is applied (see \cite{18} for more details).

\begin{table}[tbhp]
\caption{ 1D-CNN network parameters for packet detection.}
\begin{center}
\begin{tabular}{|c|c|}
\hline
Layer&Size (number of filters/neurons) \\
\hline 
Conv1D + ReLU&9\\
Conv1D + ReLU&5 (filter size is 3 samples)\\
FC + ReLU&3\\
Output (Linear)&1\\
\hline
\end{tabular}
\label{table_2}
\end{center}
\end{table}

 The same 1D-CNN architecture, whose parameters are set as in Table \ref{table_2}, is used for all experiments. Filter size of first convolution layer is chosen as a half of the STS sample-length (8 samples), and stride of 1 sample is applied (Fig. \ref{fig:1D-CNN}). Note that we do not exploit full flexibility of 1D-CNN architecture since we apply fixed number of input channels as well as the fixed-length filters. We apply such fixed architecture to make the analysis of the proposed algorithm in terms of  performances and complexity easier. We note that further optimization of the number of input channels and the input filter lengths may further improve performance vs complexity trade-off. Finally, note that the larger the length of the input block, the complexity of the first layer increases, however, the number of blocks to be processed per unit time decreases. Careful investigation of different network given is presented in Sec. IV.B.

\section{Numerical Results}

In this section, we discuss the packet detection performance of both CNN-based and conventional methods in terms of the mean absolute error (MAE) under different SNRs, while taking into account miss detection and false alarm rates. Furthermore, we investigate the computational complexity of the proposed CNN-based algorithm for packet detection for different input block lengths, and compare them to the conventional method in terms of the approximate number of floating point operations per second (FLOPS).

\textbf{Packet Detection Performance:} 
In the following experiments, the value selected for the number of input channels is set to 4 for all input block lengths. We consider MAE performance, while taking into account also the probability of miss detection and the probability of false alarm. 

%\vspace{-0.3cm}
%\begin{table}[tbhp]
%\caption{ Optimized number of 1D-CNN input channels.}
%\begin{center}
%\begin{tabular}{|c|c|c|c|c|c|c|c|}
%\hline
%Block length & 25 & 50 & 100 & 200 & 400 & 800 & 1600 \\
%\hline 
%No. of channels & 5 & 5 & 4 & 10 & 8 & 8 & 16 \\
%\hline
%\end{tabular}
%\label{table_3}
%\end{center}
%\end{table}

\begin{figure}[htbp]
 \centerline{\includegraphics[width=1\columnwidth, height=2.4in]{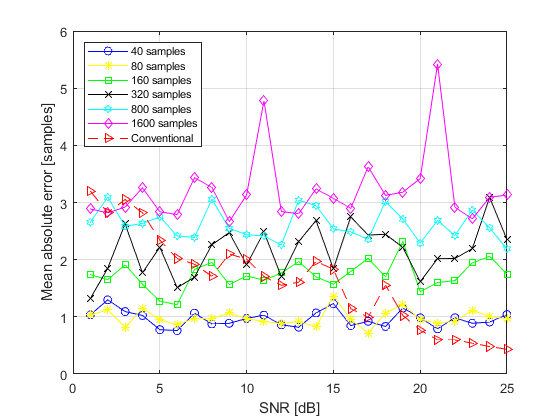}}
  \caption{MAE of 1D-CNN networks vs SNR.}
  \label{fig:MAE_O}
\end{figure}

Figure \ref{fig:MAE_O} presents MAE packet detection performance of 1D-CNN architectures as a function of the received SNR evaluated over the test set. The figure also includes the results obtained using conventional method after both coarse and fine packet start sample estimation is applied. We note that the 1D-CNN approach shows excellent robustness to the variations of SNR as compared to the conventional method, whose performance deteriorates for lower SNRs. In addition, as the input block lengths decrease, the 1D-CNN packet detector outperforms the conventional method. Although this can be, to the large extent, attributed to the fact that the estimated packet start sample value $\tau_{S}$ is bounded by the input block size (thus the estimation error naturally reduces by decreasing the input block length), we still note that 1D-CNNs taking as large as 320 samples for input blocks perform comparably with the conventional detector that slides across input blocks of 80 samples (Sec. III.A), and even outperform conventional detector for SNRs below 7 dB.

\begin{figure}[htbp]
 \centerline{\includegraphics[width=1\columnwidth, height=2.6in]{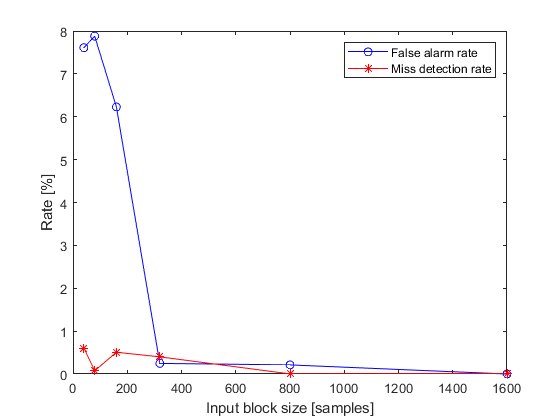}}
  \caption{ 1D-CNN miss and false detection rate for different input block sizes.}
  \label{fig:MD/FD_O}
\end{figure}

For the same setup, Figure \ref{fig:MD/FD_O} presents the miss detection and false alarm rates for different input block sizes. The results are expressed as a percentage of miss or false detected packets averaged across the entire test set (i.e., across all SNRs). For comparison, for the same testing conditions, the conventional method exhibits superb performance of miss detection rate equal $0.007 \%$ and false alarm rate equal $0.009 \%$. For 1D-CNN-based packet detectors, although the results vary across the range of input block lengths showing particularly high false alarm rates for small input block sizes, the performance gradually improves for larger input block lengths, achieving sub-$0.1\%$ miss detection and false alarm rates.

\begin{figure}[htbp]
 \centerline{\includegraphics[width=0.85\columnwidth, height=2.4in]{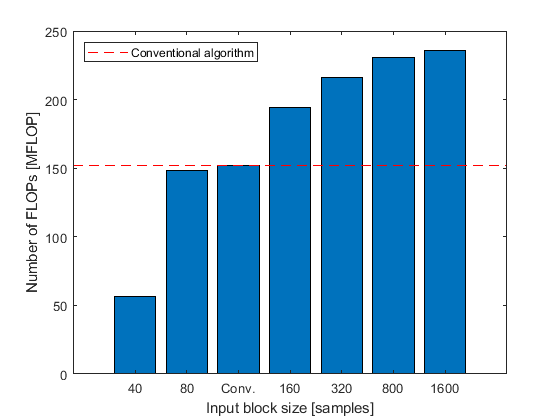}}
  \caption{Number of mega FLOPS comparison for 1D-CNNs.}
  \label{fig:FlopBar}
\end{figure}

\textbf{Computational Complexity Analysis:} Approximate number of FLOPS of 1D-CNN architectures as a function of input block lengths is presented in Fig.  \ref{fig:FlopBar}, with the conventional method included for reference. Assuming the sampling rate of 1 MHz for IEEE 802.11ah scenario used in our experiments, we present in Fig. \ref{fig:FlopBar} the number of FLOPS for 1D-CNNs with the same number of input channels. 

The complexity of each layer of 1D-CNN may be computed by calculating the number of additions and multiplications within each layer. The total number of FLOPS for a CNN depends on the input block size, however, note that although larger input blocks lead to more complex network, they also reduce the number of blocks processed per second. According to \cite{Karra_2017}, the complexity of a single convolution layer depends on filter length $F$, number of input ($ch_i$) and output ($ch_o$) channels, and output width $K$, while the complexity of FC layer is determined by the input ($N_i$) and the output ($N_o$) size. Mathematical expressions used for calculating an approximate number of FLOPS (multiplications and additions) in a single layer are given in Table \ref{table_4} \cite{Karra_2017}. 

\begin{table}[tbhp]
\caption{ Approximate layer complexity}
\begin{center}
\begin{tabular}{|c|c|}
\hline
Layer/Operation&Expression  \\
\hline 
Conv1D/ MUL&$F\ast ch_i \ast ch_o \ast K$\\

Conv1D/ADD&$F\ast (ch_i+1) \ast ch_o \ast K$\\

FC/MUL&$N_i \ast N_o$\\

FC/ADD&$(N_i+1) \ast N_o$\\
\hline
\end{tabular}
\label{table_4}
\end{center}
\end{table}

Regarding the conventional packet detection algorithm, recall that it consists of two parts: coarse and fine estimation. During the coarse estimation, it uses sample-by-sample processing of input blocks of length 80 samples. The FLOPS count for the coarse packet detection are derived by calculating the number of multiplications and additions for a single input block of length 80 samples, multiplied by the number of blocks processed per second. Complexity of the fine estimation, which is run only when the coarse estimation detects the start of the packet, is neglected. 

From Fig. \ref{fig:FlopBar}, for the same number of input channels, the complexity of 1D-CNNs are comparable with the conventional algorithm, especially for smaller input block lengths. Comparing these results with Figs. \ref{fig:MAE_O} and \ref{fig:MD/FD_O}, we note that 1D-CNNs are able to outperform conventional methods under reduced computational burden for the receiver side, while still being inferior in miss detection and false alarm rates, which calls for a proper compromise when searching for desirable performances. For example, we note that, by adapting the values of 1D-CNN parameters (number of input channels and filter input lengths), one can manage to keep the complexity of 1D-CNN below the complexity of the conventional algorithm for input blocks of lengths as large as 800 samples. Thus 1D-CNN offers wide operational range for balancing between MAE performance, computational effort in MFLOPs, miss detection and false alarm rates.    

\section{Conclusions}

In this paper, we performed in-depth investigation of 1D-CNN architectures for preamble-based packet detection in Wi-Fi. The presented results show promising performance and complexity features of 1D-CNNs compared to conventional methods. Our ongoing work is concerned with demonstrating these results using software-defined radios in real-world indoor environment.

\section{Acknowledgement}

This work has received funding from the European Union's Horizon 2020 research and innovation programme under Grant Agreement number 856967.

\end{document}